\documentclass[twocolumn,aps,prl,superscriptaddress]{revtex4-2}
\usepackage[utf8]{inputenc}
\usepackage{epsf,graphicx}
\usepackage{multirow}
\usepackage{amsmath,gensymb,amssymb}
\usepackage{lipsum}
\usepackage{dcolumn}% Align table columns on decimal point
\usepackage{bm}% bold math
\usepackage{hyperref}
\usepackage{natbib}
\usepackage{color}

\begin{document}

\title{Effects of strong turbulence for water waves}

\author{E.A. Kochurin}
\email{kochurin@iep.uran.ru}
\affiliation{Institute of Electrophysics, Ural Division, Russian Academy of Sciences, Yekaterinburg, 620016 Russia}
\affiliation{Skolkovo Institute of Science and Technology, 121205, Moscow, Russia}

\author{E.A. Kuznetsov}
\email{kuznetso@itp.ac.ru}
\affiliation{Skolkovo Institute of Science and Technology, 121205, Moscow, Russia}
\affiliation{Lebedev Physical Institute, RAS, 119991 Moscow, Russia}
\affiliation{Landau Institute for Theoretical Physics, RAS, Chernogolovka, 142432 Moscow region, Russia}
\date{\today}

\begin{abstract}
The results of direct numerical simulation of plane-symmetric turbulence of water waves for potential flows within the framework of conformal variables taking into account low-frequency pumping and high-frequency viscous dissipation are presented. In this model, for a wide range of pumping amplitudes, the weak turbulence regime was not detected. It is shown that for typical turbulence parameters, the main effects are the processes of wave breaking, the formation of cusps on wave crests, which make the main contribution to the turbulence spectra with a dependence on frequency and wavenumber with the same exponent equal to $-4$. In this strongly nonlinear regime, the probability density of wave steepness at large deviations has power-law tails responsible for the intermittency of turbulence.
\end{abstract}

\maketitle

{\bf Introduction.}
It is well known that the main mechanism of dissipation of ocean waves even with moderate wind pumping is wave breaking (see, for example, \cite{Zakharov-AGU}). As numerous observations show, the breaking process begins at the wave crest in the direction perpendicular to its front. This process is one-dimensional. At the initial stage of breaking, a singularity in the form of a cusp is formed in the wave profile. Thus, at a given point $x_0$, the surface deviation profile $y=\eta(t)$ and its derivative $d\eta(t)/dt$ remain smooth functions of time until the moment of breaking $t=t_0$. At $t=t_0$, the quantity $d\eta(t)/dt$ undergoes a jump, and its derivative will be proportional to the $\delta$-function of time. Thus, at the observation point $x=x_0$, the second derivative of $\eta(t)$ can be written as
\begin{equation} \label{sing}
	\frac{d^2 \eta}{d t^2}=\sum_i \Gamma_i \delta(t-t_i) + \mbox{regular terms},
\end{equation}
where $t_i$ are the times of singularity formation, $\Gamma_i$ is the jump in the value of $d^2 \eta/d t^2$ at the moment of singularity. Performing the Fourier transform of the singular part (\ref{sing}), we obtain the dependence of $\eta_{\omega}$ on frequency:
\[
\eta_{\omega}=\frac {1}{\sqrt{2\pi}\omega^2}\sum_i \Gamma_i e^{-i\omega t_i}.
\]
Considering the quantities $\Gamma_i$ and $t_i$ to be random, after averaging we find the contribution of singularities to the spectrum $E_{\omega}$:
\begin{equation} \label{spectrum}
	E_{\omega}=\frac{g}{2\pi T}\langle|\eta_{\omega}|^2\rangle=\frac{g\nu}{2\pi \omega^4} \langle\Gamma^2\rangle,
\end{equation}
where $g$ is the acceleration of gravity, $T$ is the averaging time, $\nu$ is the average frequency of occurrence of singularities, and the angular brackets $\langle...\rangle$ mean averaging over time. The spectrum (\ref{spectrum}) was obtained in \cite{Kuznetsov04}. It is important to note that the frequency dependence (\ref{spectrum}) proportional to $\omega^{-4}$ coincides with the Zakharov-Filonenko spectrum \cite{ZakharovFilonenko} of weak turbulence, which has now been very well reproduced both experimentally and numerically for isotropic wave propagation on water surface (see the review \cite{Onorato} and references therein).
This is purely coincidence, since the spectrum of weak turbulence implies a weakly nonlinear regime, when the nonlinearity is small compared to the dispersion of linear gravity waves on deep water
\begin{equation}\label{disp}
	\omega=\sqrt{gk}.
\end{equation}
The spectrum (\ref{spectrum}) corresponds to strong nonlinearity when the characteristic gradients of $\eta(x,t)$ are not small.

From the same ideas, in the one-dimensional formulation $d=1$, in the strong turbulence regime, the dependence of the spectrum on wavenumber $k$ has the same power-low exponent as the frequency spectrum (\ref{spectrum}): $E_k\propto k^{-4}$. As was shown in \cite{Kuznetsov04}, in the two-dimensional formulation, the spectrum on $k$ retains the same power dependence as for $d=1$: $\propto k^{-4}$. The fact that the slope of the spectra $E_{\omega}$ and $E_{k}$ is the same both in the frequency domain and in the wavenumber region implies that instead of the root dependence in the dispersion relation (\ref{disp}), there is a linear relationship $\omega\sim k$. This contradiction is explained by the fact that at a strongly nonlinear stage of evolution, the dispersion law $\omega=\omega(k)$ can deviate significantly from (\ref{disp}).

In this work, we consider the problem of surface gravity wave turbulence in deep water in a one-dimensional plane-symmetric formulation. For numerical modeling, we use a conformal transformation of the flow region into the lower half-plane \cite{DyachenkoKuznetsovSpectorZakharov, DyachenkoZakharovKuznetsov96}. The transition to conformal variables allows us to effectively identify the formation of singularities on the free surface and to study strong turbulence of surface waves. Direct numerical integration of the Euler equation for potential flows taking into account the boundary conditions on the free surface is a very complicated problem, for this reason it has not yet been completely solved.

It should be noted that in the one-dimensional formulation, the classical Zakharov-Filonenko spectrum \cite{ZakharovFilonenko} is not realized because the four-wave matrix elements, as shown in \cite{5wave}, vanish on the corresponding resonance surface. The authors of \cite{5wave} suggested that in the one-dimensional formulation, the spectra of weak turbulence should be determined by the five-wave resonant interaction. For this reason, the Kolmogorov-Zakharov spectrum for weak turbulence of plane gravity waves does not coincide with the Zakharov-Filonenko spectrum \cite{ZakharovFilonenko}, which describes isotropic two-dimensional turbulence.

In this paper, by direct numerical integration of the equations of motion in conformal variables, we show that in the strong turbulence regime, a power-law spectrum (\ref{spectrum}) is realized both in the frequency range and in the wavenumber region with the same exponent equal to $-4$, thereby confirming the predictions of the theory \cite{Kuznetsov04} . In this regime, strongly non-Gaussian power-law tails are formed in the probability density functions for gradients indicating the intermittency of strong turbulence.

Note that we have carried out numerical experiments in a wide range of small pumping amplitudes in order to detect weak turbulence spectra due to five-wave interactions. According to the predictions of \cite{5wave}, such interaction should generate a spectrum of the Kolmogorov type. Such a weakly turbulent spectrum for gravity waves is not observed. Note that the weak turbulence regime in plane geometry is realized due to five-wave interactions of capillary waves only in the region of small scales, where surface tension effects dominate, see the experimental and computational works \cite{cwt1,cwt2,cwt3}.

{\bf Basic equations and numerical scheme.}
Consider the equations of motion of an ideal fluid with a free surface for two-dimensional potential flows in a vertical gravity field in the deep water approximation. We introduce a Cartesian coordinate system $\{x,y\}$ so that the unperturbed shape of the boundary corresponds to $y=0$, and the gravity field vector is directed against the vertical axis ${\bf g} \| \hat y$. The function $y=\eta(x,t)$ describes the deviation of the free surface from the unperturbed state.
The velocity potential (${\bf v}=\nabla \phi(x,y,t)$) satisfies the Laplace equation due to the incompressibility of the fluid: $\Delta \phi=0$.
On the free surface of the fluid, the dynamic and kinematic boundary conditions are set for the Laplace equation:
\begin{equation}\label{eqx1}
	\phi_t+\frac{1}{2}|\nabla \phi|^2=-g\eta,
\end{equation}
\begin{equation}\label{eqx2}
	\eta_t+\eta_x\phi_x=\phi_y.
\end{equation}
At $y\to -\infty$ the fluid motion decays $\phi\to 0$.

As was shown by V.E.~Zakharov \cite{Zakharov1,Zakharov2}, the equations of motion (\ref{eqx1}) and (\ref{eqx2}) are the Hamiltonian type:
\begin{equation} \nonumber
	\eta_t=\delta H/\delta \psi,\qquad \psi_t=-\delta H/\delta \eta,
\end{equation}
where Hamiltonian
\begin{equation}\label{Ham}
	H=\frac{1}{2}\iint \limits_{y\leq \eta}{|\nabla \phi|^2}dxdy
	+\frac{g}{2}\int\limits_{-\infty}^{+\infty}\eta^2dx
\end{equation}
coincides with the total energy of the system, and $\psi$ is the value of the potential $\phi$ on the free surface:
\[
\psi=\phi (x,y=\eta,t).
\]
The value of $\phi$ is found from the solution of the Dirichlet problem for the Laplace equation, which is expressed through the corresponding Green's function $G$. The expression for $G$ can be obtained as an infinite series and can therefore be used within the framework of perturbation theory, in particular, in the weak turbulence theory. However, the use of conformal variables allows one to effectively study strong turbulence.

Let us introduce dimensionless variables by choosing the characteristic scale $k_0=2\pi/\lambda_0$ corresponding to long waves $k_0\ll k_{gc}=(\rho g/\sigma)^{1/2}$, where $\rho$ is the fluid density, $\sigma$ is the surface tension coefficient. The characteristic time scale corresponds to $t_0=2\pi/\omega_0=2\pi(gk_0)^{-1/2}$. Further in the work, the dimensionless quantities are used: $\tilde t=t \omega_0,$ $\tilde x =x k_0,$ $\tilde \eta =\eta k_0,$ $\tilde \phi =\phi k_0^2/\omega_0$. For convenience, the tilde signs are omitted below.

In conformal variables, the system of equations (\ref{eqx1}) and (\ref{eqx2}) has the form:
\begin{equation}\label{eq1}
	Y_t=\left(Y_u\hat H-X_u\right)\frac{\hat H \Psi_u}{J}- \hat \gamma_k Y,
\end{equation}
\begin{equation}\label{eq2}\Psi_t=\frac{(\hat H \Psi_u)^2-\Psi_u^2}{2J}+\hat H\left(\frac{\hat H \Psi_u}{J}\right)\Psi_u-Y+\mathcal{F}(\textbf{k},t)- \hat \gamma_k \Psi,
\end{equation}
where $z=X(u,v)+iY(u,v)$ maps the region occupied by the fluid into the lower half-plane $v\leq 0$. The free boundary corresponds to the line
$v=0$. In these equations, $X$ and $Y$ are the values of $x$ and $y$ at $v=0$, the shape of the fluid and the velocity potential are specified parametrically, i.e., $\eta=Y(X)$ and $\psi=\Psi(X)$;
$J=X_u^2+Y_u^2$ is the Jacobian of the transformation, $\hat H$ is the Hilbert transform, which for periodic functions has the form: $\hat H f=(2\pi)^{-1}\mbox{p.v.}\int_0^{2\pi}f(x')\cot[(x-x')/2]dx'$. In the Fourier representation, $\hat H$ is the multiplication operator $\hat H =i\mbox{sign}(k)$.

To model wave turbulence, we introduce additional terms responsible for energy dissipation in the short-wave region and pumping concentrated in the long-wave region. The viscous dissipation operator is defined in $k$-space as
\begin{equation*}\label{diss}
	\hat \gamma _{k} f_k =\gamma _{0} k^2 f_k,
\end{equation*}
where $\gamma _{0}$ is assumed to be constant for $k<k_d$. In order to stabilize small-scale instability and to reduce the wave breaking effect, which leas to form mutually ambiguous regions of the boundary shape, i.e., angles greater than $\pi/2$, the amplitude of $\gamma _{0}$ for $k\geq k_d$ is increased by three orders of magnitude.

The pumping term $\mathcal{F}(\textbf{k},t)$ is defined in Fourier space as follows:
\begin{equation*}
	\mathcal{F}(\textbf{k},t) =F(k)\cdot \exp [iR(\textbf{k},t)],
\end{equation*}
\begin{equation*}
	F(k) =F_{0}\cdot \exp [-L_0^4(k-k_{f})^{4}].
\end{equation*}
Here $R(\mathbf{k,}t)$ are random numbers uniformly distributed in the interval $[0,2\pi ]$, $F_{0}$ is a constant, $k_f$ is the wavenumber at which the pumping amplitude reaches its maximum, and $L_0$ specifies the pumping width.

The integration of the equations of motion (\ref{eq1}) and (\ref{eq2}) over time is performed by the explicit fourth-order Runge-Kutta method with a step $dt$. All calculations are performed in a periodic domain of length $L=2\pi$ with the following parameters: $dt=2.5\cdot 10^{-5}$, with the number of spatial points  $N=32\, 768$, $\gamma_0=10^{ -5}$, $k_d=5000$, $F_0=0.6\cdot 10^5$, $k_f=3$, $L_0=0.64$. To suppress the aliasing effect, a low-pass filter is used at each time integration step, eliminating higher harmonics $k\geq N/3$.

\begin{figure}[t!]
	\centering
	\includegraphics[width=1.0\linewidth]{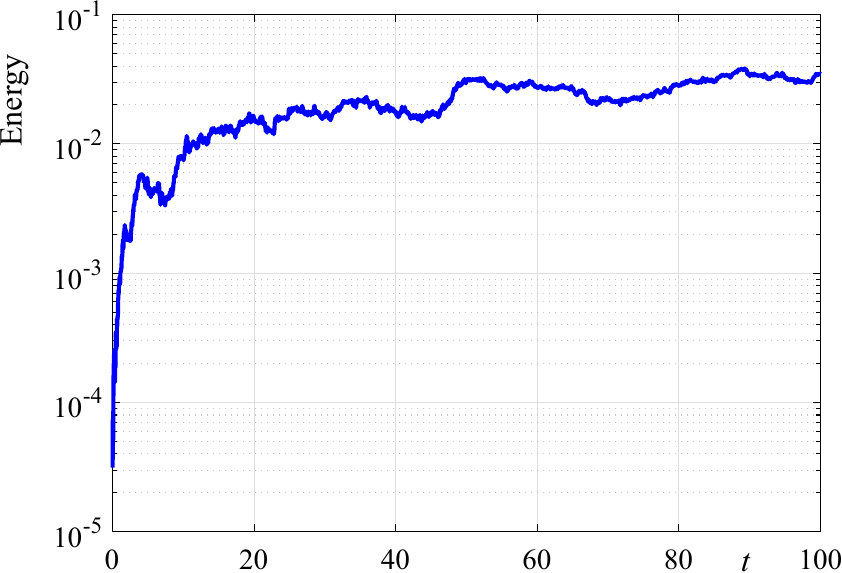}
	\caption{(Color online) Time evolution of the total energy of the system (\ref{Ham}).}
	\label{fig1}
\end{figure}

\begin{figure}[t]
	\centering
	\includegraphics[width=1\linewidth]{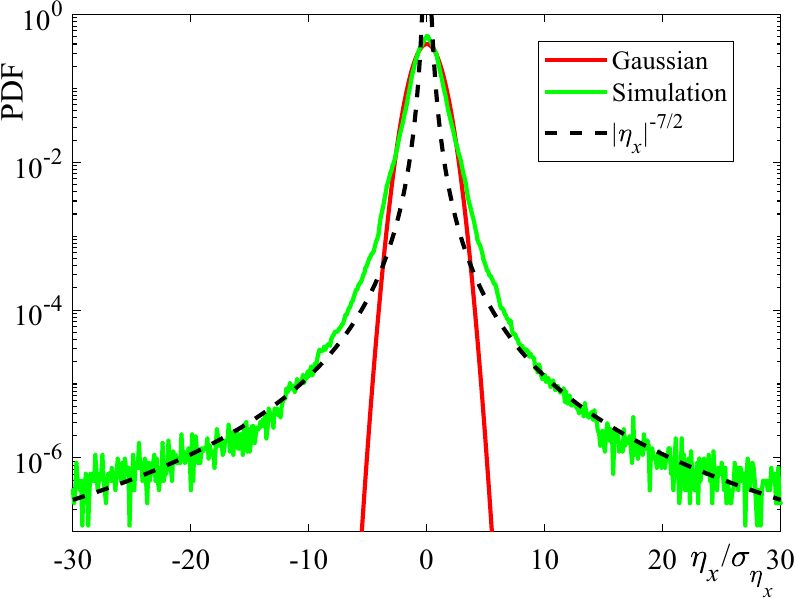}
	\caption{ (Color online) Computed probability density function for the steepness of the boundary $\eta_x$, measured relative to the standard deviation $\sigma_{\eta_x}$, in the quasi-steady state.}
	\label{fig2}
\end{figure}

\begin{figure*}[t]
	\centering
	\includegraphics[width=1\linewidth]{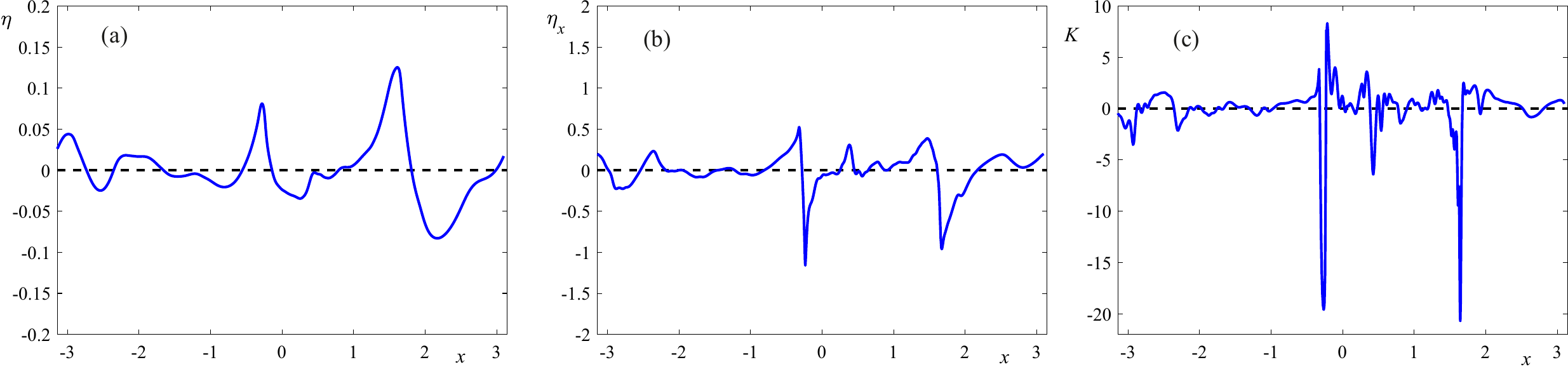}
	\caption{ (Color online) (a) Shape of the liquid boundary, (b) steepness and (c) curvature of the boundary at some moment of the quasi-steady state $t=40$.}
	\label{fig3}
\end{figure*}

\begin{figure*}[t]
	\centering
	\includegraphics[width=1.0\linewidth]{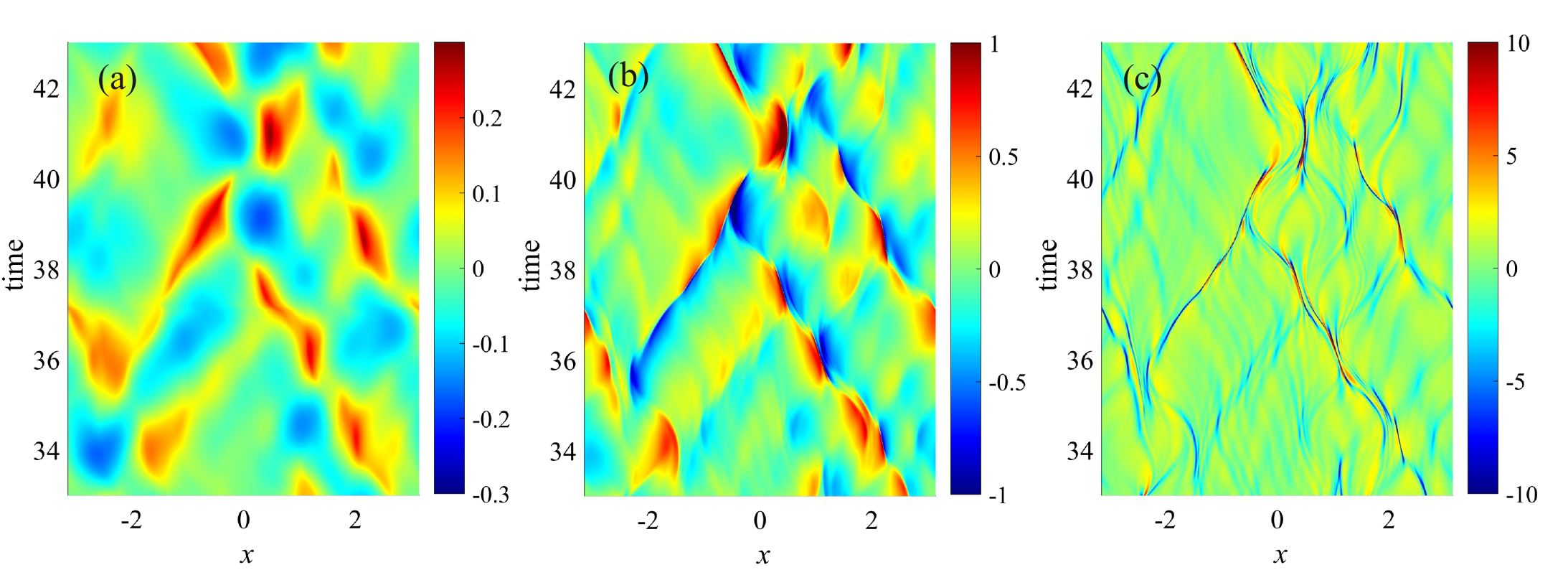}
	\caption{(Color online) Evolution of the shape (a), steepness (b) and curvature (c) of the fluid boundary in the time interval $t\in(33, 43)$.}
	\label{fig4}
\end{figure*}

{\bf Simulation results.}
In the absence of dissipation, equations (\ref{eq1}) and (\ref{eq2}) describe the process of wave collapse leading to the formation of a discontinuity in the boundary curvature in a finite time. This process has been studied quite well, see, for example, \cite{sing1,sing2,sing3}. The goal of this work is to study the quasi-stationary regime, in which dissipative effects prevent the formation of singularities. Fig.~\ref{fig1} shows the evolution of the total energy of the system for the chosen calculation parameters. Indeed, it is seen that the energy rapidly increases to a certain level and then performs rather complex oscillations. In the quasi-stationary regime, the probability density functions deviate very strongly from the Gaussian distribution, see Fig.~\ref{fig2}. In the figure, we observe long power-law tails with an exponent of $-7/2$, which indicate intermittency.

Fig.~\ref{fig3} shows the boundary shape, its steepness $\eta_x$, and surface curvature $K=\eta_{xx}/(1+\eta_x^2)^{3/2}$ at some moment in the system evolution. It can be seen that regions with high boundary curvature have formed on the liquid surface. The slope angles (steepness $\eta_x$) undergo a discontinuity at these points. Fig.~\ref{fig4} shows the time segment of the evolution of the liquid surface shape, as well as the values of $\eta_x$ and $K$ in the turbulent regime. It is evident that several shock fronts (spatially narrow regions in which the slope angles undergo a discontinuity) have formed on the liquid boundary. It is important that the shock fronts move at an almost constant velocity. This is especially evident in Fig.~\ref{fig4}~(b)~and~(c).

\begin{figure}[t]
	\centering
	\includegraphics[width=0.9\linewidth]{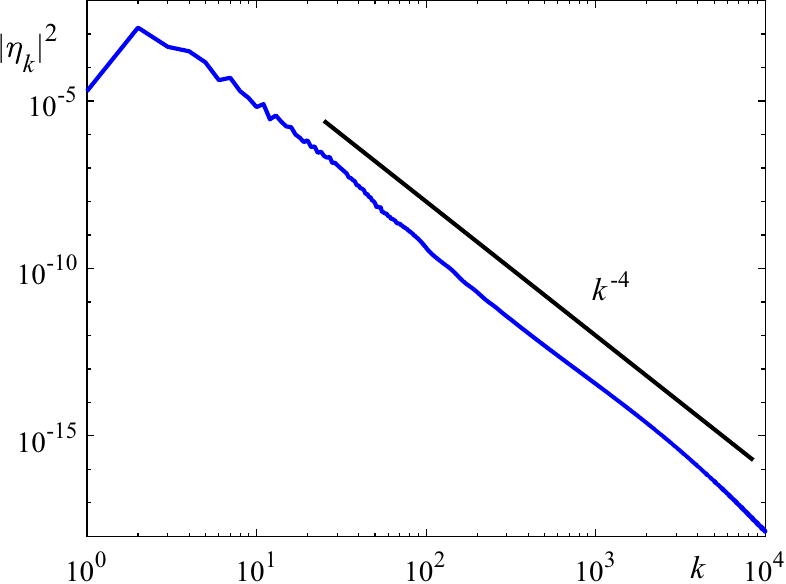}
	\caption{(Color online) Spatial spectrum of surface perturbations compared to the spectrum of $k^{-4}$.}
	\label{fig5}
\end{figure}

\begin{figure}[t]
	\centering
	\includegraphics[width=0.9\linewidth]{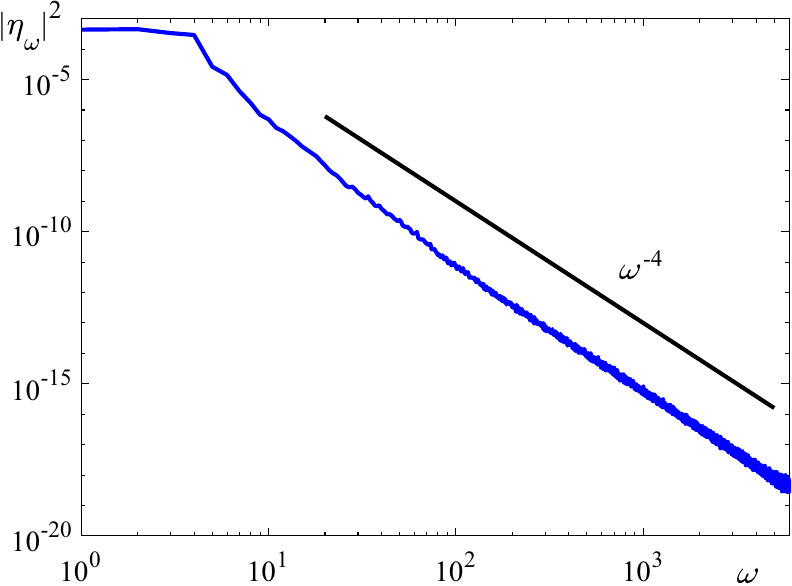}
	\caption{(Color online) Frequency spectrum of surface perturbations compared to the spectrum of $\omega^{-4}$.}
	\label{fig6}
\end{figure}

{\bf Turbulence spectra.}
The main question is what turbulence spectrum does the system demonstrate in a strongly nonlinear motion regime. Figures~\ref{fig5}~and~\ref{fig6} show the spatial and frequency spectrum of surface disturbances in the turbulent motion regime. Note that the figures show the spectra of the reconstructed function $\eta(x,t)$, not the function $Y(u,t)$ defined in conformal coordinates. The procedure for reconstructing the explicit dependence $\eta(x,t)$ is described in \cite{cwt2}. The spectrum~(\ref{spectrum}) is indeed well reproduced in our numerical experiment. The inertial interval was almost three decades in both $k$ and $\omega$ domains, which is quite large for direct numerical simulation methods. It is important that the strong turbulence spectrum (\ref{spectrum}) is also observed in direct laboratory experiments \cite{exp1,exp2,exp3} (see also the review by \cite{Onorato}).

It should be noted that the turbulence spectrum observed in Fig.~\ref{fig5}~and~\ref{fig6} differs significantly from the Phillips spectrum originally obtained based on the dimensional analysis \cite{phil}. In fact, the Phillips spectrum corresponds to the situation when the nonlinearity effects are comparable with the linear dispersion of waves (the so-called critical balance, see \cite{CB1, CB2, CB3}). In the studied regime, nonlinear effects (collapse tendencies) play a dominant role. Note also that the Kolmogorov-Zakharov spectrum of weak turbulence proposed in \cite{5wave} to describe plane-symmetric gravity waves cannot be reproduced within the framework of the numerical model used. Calculations show that the system always passes into the strong nonlinearity regime. Perhaps this fact is explained by the instability of the weak turbulence spectrum \cite{5wave} under plane symmetry conditions.

\begin{figure}[t]
	\centering
	\includegraphics[width=1\linewidth]{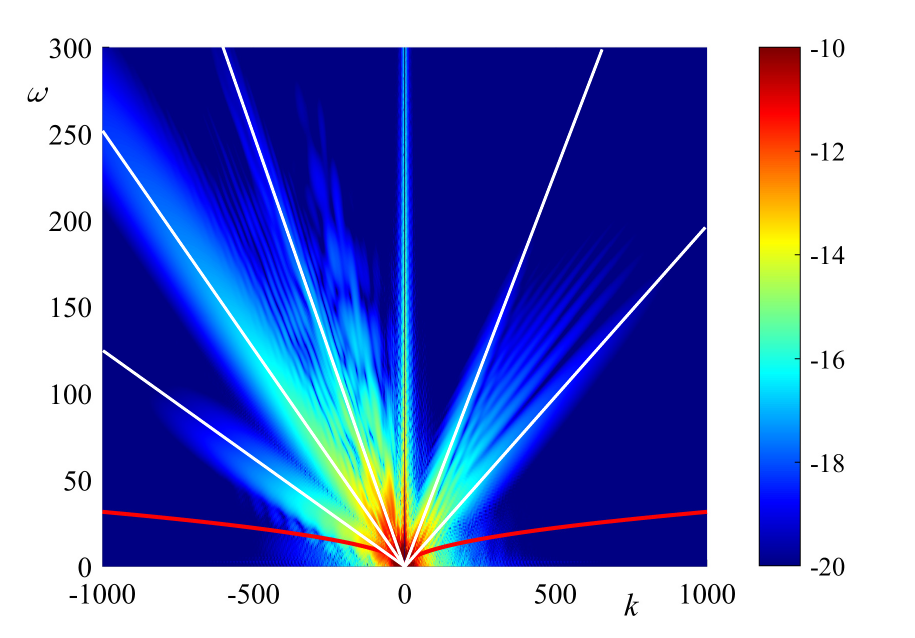}
	\caption{(Color online) Spatiotemporal spectrum of surface perturbations $\log|\eta(\omega,k)|$, red solid lines is the linear dispersion law $\omega= k^{1/2}$, white solid lines correspond to non-dispersive propagation of nonlinear waves $\omega\sim k$.}
	\label{fig7}
\end{figure}

{\bf Spatiotemporal analysis of turbulence.}
As noted earlier, the spectrum of strong turbulence (\ref{spectrum}) is valid for disturbances with frequencies and wavenumbers linearly related $\omega\sim k$. At the same time, such a dependence obviously contradicts the linear dispersion law (\ref{disp}).
In order to find out how the wave frequency can be related to the wavenumber in a strongly nonlinear motion regime, we performed a spatiotemporal Fourier analysis. Fig.~\ref{fig7} shows the spatiotemporal Fourier spectrum of the surface $|\eta(\omega,k)|$, calculated in the regime of developed turbulence. In the figure, we see that the surface disturbances are grouped along straight lines describing non-dispersive wave propagation. The observed perturbations deviate very strongly from the linear dispersion law, shown by the red solid line. In fact, we find a discrete set of shock fronts propagating with nearly constant velocity, at least for some finite time interval. Apparently, the motion with constant velocity arises from the balance of nonlinear effects and energy dissipation. Note that a similar picture was observed in \cite{Dyachenko08} in the context of searching for integrable structures on the surface of a deep water.

{\bf Concluding remarks.}
Thus, a direct numerical experiment for one-dimensional gravity wave turbulence in deep water has demonstrated that the main contribution to turbulence is due to wave breaking, which produces a power-law spectrum (\ref{spectrum}) in both the frequency range and the wavenumber space. To our knowledge, our results represent the first direct numerical confirmation of the strong turbulence spectra predicted in \cite{Kuznetsov04}. The calculated probability density function for the surface gradient in strong turbulence has power-law tails with exponents close to $-7/2$. It is interesting to note that such behavior is typical for one-dimensional Burgers turbulence \cite{pdf1,pdf2,pdf3,pdf4,pdf5}, and is also observed for strong three-dimensional sound turbulence \cite{KochurinKuznetsov}.
Note that wave collapse plays an important role in one-dimensional models of wave turbulence (the so-called MMT model \cite{MMT1}). In the MMT model, collapses lead to strong intermittency and destruction of weakly turbulent Kolmogorov-Zakharov spectra \cite{MMT2,MMT3,MMT4,MMT5,MMT6}. A similar shock-wave regime of turbulence is also realized for weakly dispersive magnetohydrodynamic waves at fluid boundaries \cite{MHD1,MHD2}.

\section{Acknowledgments}
The authors are grateful to M. Onorato, who drew our attention to the work of \cite{Onorato}.

\section{Funding}
The work was carried out within the framework of the grant of the Russian Science Foundation No. 19-72-30028.

\section{Conflict of interest}
The authors of this work declare that they have no conflict of interest.

\end{document}